\documentclass[twocolumn,showpacs,preprintnumbers,amsmath,amssymb,floatfix]{revtex4}

\usepackage{graphicx}
\usepackage{dcolumn} 
\usepackage{bm}      
\usepackage{times}
\usepackage{color}

\newcommand{\vB}{\mathbf{B}}
\newcommand{\vA}{\mathbf{A}}
\newcommand{\vv}{\mathbf{v}}
\newcommand{\vu}{\mathbf{u}}
\newcommand{\vj}{\mathbf{j}}
\newcommand{\vx}{\mathbf{x}}
\newcommand{\vk}{\mathbf{k}}
\newcommand{\vomega}{\mbox{\boldmath $\omega$}}

\begin{document}

\title{Numerical solutions of the three-dimensional 
       magnetohydrodynamic alpha-model}

\author{Pablo D. Mininni$^1$, David C. Montgomery$^2$ and 
        Annick Pouquet$^1$}

\affiliation{$^1$ Advanced Study Program, National Center for Atmospheric 
Research, P.O. Box 3000, Boulder, Colorado 80307}
\affiliation{$^2$ Dept. of Physics and Astronomy,
Dartmouth College, Hanover, NH 03755} 

\date{\today}

\begin{abstract}
We present direct numerical simulations and $\alpha$-model simulations 
of four familiar three-dimensional magnetohydrodynamic (MHD) turbulence 
effects: selective decay, dynamic alignment, inverse cascade of magnetic 
helicity, and the helical dynamo effect. The MHD $\alpha$-model is shown 
to capture the long-wavelength spectra in all these problems, allowing for 
a significant reduction of computer time and memory at the same kinetic 
and magnetic Reynolds numbers. In the helical dynamo, not only does the 
$\alpha$-model correctly reproduce the growth rate of magnetic energy 
during the kinematic regime, but it also captures the nonlinear saturation 
level and the late generation of a large scale magnetic field by the 
helical turbulence.
\end{abstract}

\maketitle

\section{\label{sec:intro}INTRODUCTION}

The ``alpha model,'' as it has come to be called in fluid mechanics, 
is a procedure whereby, by suppressing small spatial scales in a computation 
in a way that intends to do minimum damage to the accuracy with which the 
long wavelength spectral components are calculated, one can realize 
substantial savings in computing time
\cite{Holm98a,Holm98b,Chen98,Chen99a,Chen99b,Chen99c,Foias01,
Nadiga01,Holm02,Holm02b,Foias02,Ilyin03,Mohseni03}. 
In a previous paper \cite{Mininni04}, we gave a simple way to extend the 
alpha model to magnetohydrodynamics (see also \cite{Holm02,Holm02b} for 
extensions in the non-dissipative case), we specialized it to two 
dimensions, and numerically tested its predictions in a series of 
computations. These were chosen as situations where direct numerical 
simulations (DNS) that started from identical initial conditions were 
feasible. The intent of this present paper is to present comparisons 
of the same kind for three-dimensional (3D) magnetohydrodynamics (MHD). 
This is a straightforward program in the light of the two-dimensional 
(2D) investigations \cite{Mininni04} and we will draw heavily on the 
material in that paper to avoid repetition, but in 3D, new phenomena 
arise, such as the generation of magnetic fields through stretching 
by velocity gradients, and furthermore 3D is computationally more 
demanding than 2D.

In Section \ref{sec:eq}, we take the 3D alpha model MHD equations 
\cite{Mininni04} and describe briefly four problems upon which they 
will be tested against DNS treatments of the same problems. They 
are selective decay, dynamic alignment, the inverse cascade of magnetic 
helicity, and the mechanically driven turbulent dynamo. The first two 
have already been tested in 2D \cite{Mininni04} and the third has a 2D 
analogue in the inverse cascade of magnetic vector potential 
\cite{Mininni04}. The fourth is also not an unfamiliar effect, and we 
have recently been involved in addressing it for the special case of 
low magnetic Prandtl number \cite{Ponty04} and for non-helical flows.

Our conclusions reached in Secs. \ref{sec:decay}-\ref{sec:dynamo} are 
consistent for the most part with those reached for 2D MHD: the alpha 
model does a good job of reproducing the spectral behavior of the 
long-wavelength Fourier amplitudes (wavenumber $k \lesssim \alpha^{-1}$, 
where $\alpha$ is the spatial scale over which the velocity field and 
magnetic field are filtered). Because of the relative lack of surprises 
in the selective decay and dynamic alignment Sections, we rely on 
relatively brief presentations, to then focus in the study of the 
inverse cascade of magnetic helicity and the dynamo effect. Finally, 
we summarize the results in Sec. \ref{sec:sum}.

\section{\label{sec:eq}RELEVANT EQUATIONS; PROBLEMS CONSIDERED}

In familiar ``Alfv\'enic'' dimensionless units, the original MHD 
equations are
\begin{eqnarray}
\frac{\partial \vv}{\partial t} + \vv \cdot \nabla \vv &=& -\nabla {\cal P} 
     + \vj \times \vB - \nu \nabla \times \vomega , \label{eq:NS} \\
\frac{\partial \vB}{\partial t} + \vv \cdot \nabla \vB &=& 
     \vB \cdot \nabla \vv - \eta \nabla \times \vj , \label{eq:ind}
\end{eqnarray}
together with $\nabla \cdot \vv = 0 = \nabla \cdot \vB $.

The velocity field is $\vv$, the magnetic field is $\vB = \nabla \times \vA$, 
where $\vA$ is the vector potential. The electric current density is 
$\vj = \nabla \times \vB$ and the vorticity is $\vomega = \nabla \times \vv$. 
The dimensionless pressure, normalized to the (uniform) mass density is 
${\cal P}$, and is obtained by taking the divergence of Eq. (\ref{eq:NS}), 
using the incompressibility condition  $\nabla \cdot \vv = 0 $, and 
solving the resulting Poisson equation. Removing a curl from Eq. 
(\ref{eq:ind}) gives
\begin{equation}
\frac{\partial \vA}{\partial t} = \vv \times \vB - \eta \vj - \nabla \Phi ,
\label{eq:vecpot}
\end{equation}
where $\Phi$ is the scalar potential, obtainable also from a Poisson 
equation by imposing the Coulomb gauge $\nabla \cdot \vA = 0$. The 
kinematic viscosity is $\nu$ and the magnetic diffusivity is $\eta$. 
In these dimensionless units, $\nu^{-1}$ can be interpreted as a 
Reynolds number $R_e = UL/\nu$ where in laboratory (c.g.s) units, $U$ is 
a mean flow speed and $L$ is a length characteristic of it. Similarly, 
$\eta^{-1}$ can be interpreted as a magnetic Reynolds number 
$R_m = UL/\eta$. The magnetic Prandtl number is 
$P_m = \nu/\eta = R_m/R_e$.

In the alpha model the fields $\vv$ and $\vB$ are smoothed but $\vomega$ 
and $\vj$ are not \cite{Mininni04,Montgomery02}. The prescription is 
\begin{eqnarray}
\vu_s &=& \int{ \textrm{d}^3x' \, \frac{\exp \left[ -|\vx-\vx'|/\alpha 
    \right]} {4 \pi \alpha^2 |\vx-\vx'|} \vv(\vx',t)} \label{eq:us} \\
\vB_s &=& \int{ \textrm{d}^3x' \, \frac{\exp \left[ -|\vx-\vx'|/\alpha 
    \right]} {4 \pi \alpha^2 |\vx-\vx'|} \vB(\vx',t)} \label{eq:Bs}.
\end{eqnarray}
Here $\alpha$ is an arbitrary filtering length, generally chosen smaller 
than the length scales one wishes to resolve. If $\vv$ and $\vB$ are 
Fourier-decomposed
\begin{eqnarray}
\vv(x,t) &=& \int{ \textrm{d}^3k \; \vv_\vk(t) e^{i \vk \cdot \vx}} \\
\vB(x,t) &=& \int{ \textrm{d}^3k \; \vB_\vk(t) e^{i \vk \cdot \vx}} , 
\end{eqnarray}
the connection between the Fourier transforms of the smoothed fields 
$\vu_s$ and $\vB_s$ and $\vv_\vk(t)$, $\vB_\vk(t)$ are
\begin{eqnarray}
\vu_s(\vk,t) &=& \vv_\vk(t)/(1+k^2 \alpha^2) \\
\vB_s(\vk,t) &=& \vB_\vk(t)/(1+k^2 \alpha^2) , 
\end{eqnarray}
or in configuration space
\begin{eqnarray}
\vv &=& \left(1 - \alpha^2 \nabla^2\right) \vu_s \\
\vB &=& \left(1 - \alpha^2 \nabla^2\right) \vB_s . 
\end{eqnarray}
Note that we choose to smooth both the velocity and the magnetic field
 at the same length-scale, a choice appropriate for the unit magnetic 
Prandtl number ($\nu=\eta$) cases treated in this paper (for a different 
choice, see \cite{Ponty04}).

The dynamics of the alpha model \cite{Mininni04} amount to solving the 
pair,
\begin{eqnarray}
\frac{\partial \vv}{\partial t} + \vu_s \cdot \nabla \vv &=& 
     - v_j \nabla u_s^j -\nabla \widetilde{{\cal P}} + \vj \times \vB_s 
     \nonumber \\
     {} && - \nu \nabla \times \vomega , \label{eq:alpNS} \\
\frac{\partial \vB_s}{\partial t} + \vu_s \cdot \nabla \vB_s &=& 
     \vB_s \cdot \nabla \vu_s - \eta \nabla \times \vj , \label{eq:alpind}
\end{eqnarray}
where it is to be emphasized that in Eqs. (\ref{eq:alpNS},\ref{eq:alpind}), 
$\vv$, $\vj$, and $\vomega$ are not smoothed. $\widetilde{{\cal P}}$ is 
to be determined, as before, from the relevant Poisson equation.

In rectangular periodic boundary conditions (which we employ throughout), 
the ideal ($\nu = 0 = \eta$) invariants that have been identified for 
Eqs. (\ref{eq:alpNS},\ref{eq:alpind}) are the energy $E$
\begin{equation}
E = \frac{1}{2} \int{ \left(\vu_s \cdot \vv + \vB \cdot \vB_s \right) 
    \textrm{d}^3x } ,
\label{eq:E}
\end{equation}
the cross helicity $H_C$,
\begin{equation}
H_C = \frac{1}{2} \int{ \vv \cdot \vB_s \, \textrm{d}^3x } ,
\label{eq:Hc}
\end{equation}
and the magnetic helicity $H_M$,
\begin{equation}
H_M = \frac{1}{2} \int{ \vA_s \cdot \vB_s \, \textrm{d}^3x } .
\label{eq:Hm}
\end{equation}

In the presence of non-zero $\eta$ and $\nu$, the decay rates for $E$, 
$H_C$, and $H_M$ can readily been shown to be
\begin{eqnarray}
\frac{\textrm{d} E}{\textrm{d} t} &=& -\nu \int{ \vomega_s \cdot \vomega
    \, \textrm{d}^3x } - \eta \int{j^2 \, \textrm{d}^3x } \label{eq:Edis} \\
\frac{\textrm{d} H_C}{\textrm{d} t} &=& - \frac{1}{2} \nu 
    \int{ \vomega \cdot \vj_s \, \textrm{d}^3x } - \frac{1}{2} \eta 
    \int{ \vomega \cdot \vj \, \textrm{d}^3x } \label{eq:Hcdis} \\
\frac{\textrm{d} H_M}{\textrm{d} t} &=& - \eta 
    \int{ \vj \cdot \vB_s \, \textrm{d}^3x } \label{eq:Hmdis}
\end{eqnarray}

When we write $\vomega_s$ or $\vA_s$, we mean that the same smoothing 
recipe as in Eqs. (\ref{eq:us},\ref{eq:Bs}) has been applied to the 
unsmoothed fields $\vomega$ or $\vA$. It is possible, and sometimes 
desirable, to use different smoothing lengths $\alpha_v$, $\alpha_B$ 
for the mechanical and magnetic quantities \cite{Mininni04,Ponty04}.

$E$, $H_C$ and $H_M$ as defined here are the ideal invariants of 
Eqs. (\ref{eq:alpNS},\ref{eq:alpind}), and reduce, as $\alpha \to 0$, to the 
usual ideal 3D MHD invariants. Sometimes, to make the global quantities 
agree at $t=0$ for initial-value runs, we may initially normalize the 
Fourier coefficients of the fields, by multiplication by a common 
factor, to bring $E$ and $H_C$ into exact initial agreement with 
the corresponding numbers for the ideal 3D MHD invariants (note that 
$H_M$ involves two smoothed fields, and therefore can not be matched to 
the DNS initial conditions at the same time). Hereafter, such global 
quantities as $E$, $H_C$, etc., will be referred to unit volume.

It is well known that for decaying turbulent situations, the presence 
of enough initial $H_M$ or $H_C$ can lead to a late-time state in 
which the ratios $|H_M/E|$ or $|H_C/E|$ can be close to maximal. The 
first situation, called ``selective decay,''  
\cite{Matthaeus80,Ting86,Kinney95} leads to a late-time quasi-steady 
state in which the remaining energy is nearly all magnetic and is 
nearly all condensed into the longest wavelength modes allowed 
by the boundary conditions. The second situation, called 
``dynamic alignment,'' \cite{Grappin83,Pouquet86,Ghosh88} 
leads to a late-time quasi-steady state in which $\vv$ and $\vB$ are 
nearly parallel or anti-parallel. In both cases, the states can be 
very long-lived because the nonlinear transfer to small scales has 
essentially been shut down (``suppression of nonlinearity''). We 
illustrate these two situations in Secs. \ref{sec:decay} and 
\ref{sec:align}.

Inverse cascade processes \cite{Lilly69,Mazure75,Pouquet76,Hossain83} 
are those wherein excitations externally injected at the small scales 
are preferentially transferred to the larger scales and pile up there, 
creating coherent macroscopic structures at large scales where none 
were present initially. A quantity which can be inversely cascaded 
in 3D MHD is $H_M$ \cite{Mazure75,Pouquet76}. We illustrate this with 
an externally-driven run in Sec. \ref{sec:inverse}.

Dynamo processes (see Ref. \cite{Brandenburg04} for a review) are those 
whereby mechanical injection of excitations transfer energy to 
magnetic fields, causing them to amplify. A novel example of helical 
dynamo action using the alpha-model is treated in Sec. \ref{sec:dynamo}.

In all four cases, well-resolved DNS solutions are regarded as baseline 
truths against which alpha-model computations are to be tested.

\section{\label{sec:decay}SELECTIVE DECAY}

In selective decays, energy decays rapidly relative to magnetic helicity, 
if any \cite{Matthaeus80,Ting86,Kinney95}. In order to display the 
process most clearly, it helps to start an initial-value decay run with 
a significant amount of magnetic helicity. One way to accomplish this 
is to make the initial values of $\vv$ and $\vB$ out of what are 
called ``ABC'' flows. We define
\begin{eqnarray}
\vv_{ABC} &=& \left[B \cos(ky) + C \sin(kz) \right] \hat{x} + \nonumber \\
    {} && + \left[A \sin(kx) + C \cos(kz) \right] \hat{y} + \nonumber \\
    {} && + \left[A \cos(kx) + B \sin(ky) \right] \hat{z} \label{eq:ABC}
\end{eqnarray}
for arbitrary real numbers $A$, $B$, $C$, and $k$. $\vv_{ABC}$ is an 
eigenfunction of the curl. The specific initial conditions chosen are
\begin{eqnarray}
\vv(t=0) = \sum_{k=k_{bot}}^{k_{top}} v_0 \left[ \vv_{ABC} (k,\phi_k) + 
    \hat{\vv}(\vk) e^{i \vk \cdot \vx} \right] \\
\vB(t=0) = \sum_{k=k_{bot}}^{k_{top}} b_0 \left[ \vv_{ABC} (k,\phi_k) + 
    \hat{\vB}(\vk) e^{i \vk \cdot \vx} \right] .
\end{eqnarray}
The notation $ \vv_{ABC} (k,\phi_k)$ means that for each $\vk$ in the 
summation, a random phase $\phi_k$ is added to the arguments of the 
sines and cosines for that $k$. The summations are over all the $\vk$ 
values (which lie on a lattice in $\vk$ space defined by the periodic 
boundary conditions) between radii $k_{bot}$ and $k_{top}$. The 
$\hat{\vv}(\vk)$ and $\hat{\vB}(\vk)$ represent added random 
perturbations.

Here, we have chosen $A=B=C=1$, $k_{bot}=6$, $k_{top}=10$, and 
$v_0$, $b_0$ are chosen to make the initial 
$\left< v^2 \right> = \left< B^2 \right> = 1$, where ``$\left< . \right>$'' 
means a spatial average over the basic box. It is also the case that 
initially, $\left< \vv \cdot \vB \right> = 0$. Random modes 
$\hat{\vv}(\vk)$ and $\hat{\vB}(\vk)$ are added with an energetic level 
to initially give 
$\left< \vA \cdot \vB \right> = 0.5 \left<|\vA| |\vB| \right>$. 
The dimensionless inverse Reynolds numbers are $\nu=\eta=0.002$.

\begin{figure}
\includegraphics[width=9cm]{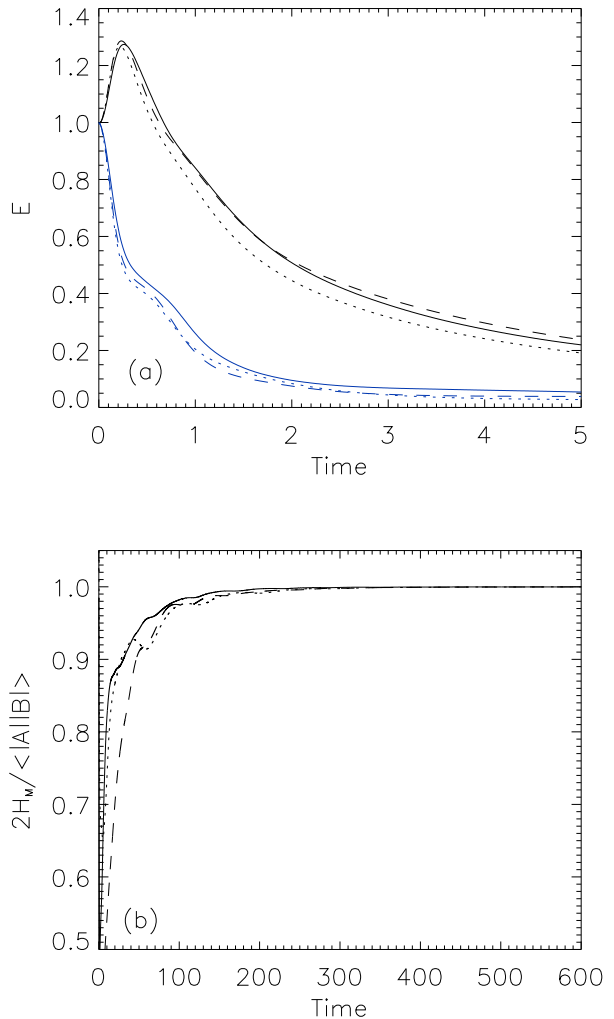}
\caption{(color online) (a) Magnetic energy (upper curves) and kinetic 
     energy (lower blue curves) as a function of time until $t=5$, and 
     (b) relative magnetic helicity as a function of time until $t=600$, 
     for the selective decay runs. Solid lines correspond to DNS, dashed 
     lines to $128^3$ $\alpha$-model simulations, and dotted lines to 
     $64^3$ $\alpha$-model simulations.}
\label{fig:decay_E}
\end{figure}

Three runs for a typical case are displayed. The first of these is a 
well-resolved DNS run at a resolution of $256^3$, with de-aliasing 
achieved by zeroing out all Fourier coefficients with $k>256/3$, a 
method that will be used throughout (usually referred to as the 
``$2/3$ rule''). Then two $\alpha$-model runs are performed with the 
same initial conditions, a $128^3$ run with $\alpha=1/20$ and a 
$64^3$ run with $\alpha=1/10$. The same values of $\nu$, $\eta$ apply 
to all three runs. The caption of Fig. \ref{fig:decay_E}.a identifies 
the decaying energies (kinetic energy $E_K$ and magnetic energy $E_M$) 
as functions of time. Fig. \ref{fig:decay_E}.b shows the ratio 
$\left<\vA \cdot \vB \right>/\left<|\vA||\vB|\right>$ as a function 
of time for the three runs; it has increased to above $0.999$ by the 
final time.

\begin{figure}
\includegraphics[width=9cm]{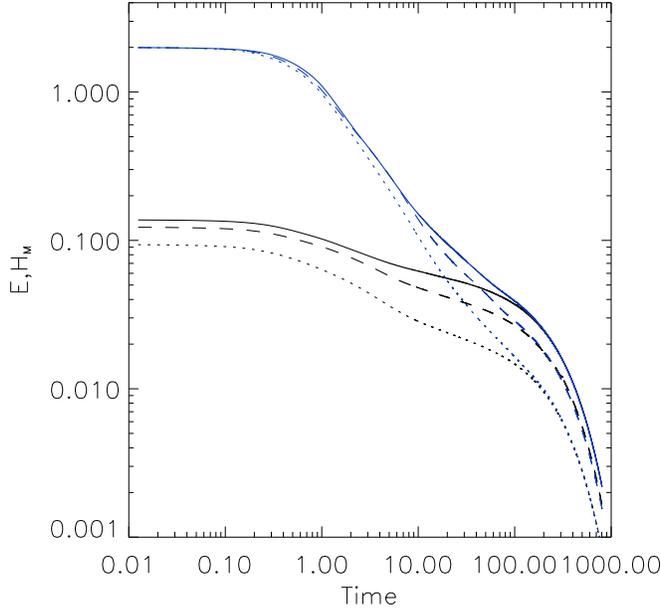}
\caption{(color online) Total energy (upper blue curves) and magnetic 
     helicity (lower curves) as a function of time. Labels are as in Fig. 
     \ref{fig:decay_E}.}
\label{fig:decay_Elog}
\end{figure}

Fig. \ref{fig:decay_Elog} shows the (unnormalized) energies and 
magnetic helicities for the three runs. Note that by normalizing 
the DNS and $\alpha$-model initial conditions to have equal energies, 
it has meant that the $\alpha$-model magnetic helicities have necessarily 
started at lower initial values than those of the DNS.

\begin{figure}
\includegraphics[width=9cm]{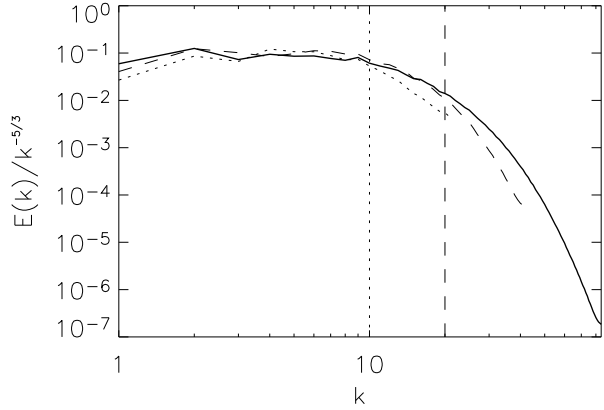}
\caption{Total energy spectrum compensated by Kolmogorov's $-5/3$ law, 
    for the three dynamic alignment runs (labels are as in Fig. 
    \ref{fig:decay_E}), at $t=10$. Vertical dotted and dashed lines 
    indicate respectively the scales $\alpha^{-1} = 10$ and $20$.}
\label{fig:decay_earlyspec}
\end{figure}

\begin{figure}
\includegraphics[width=9cm]{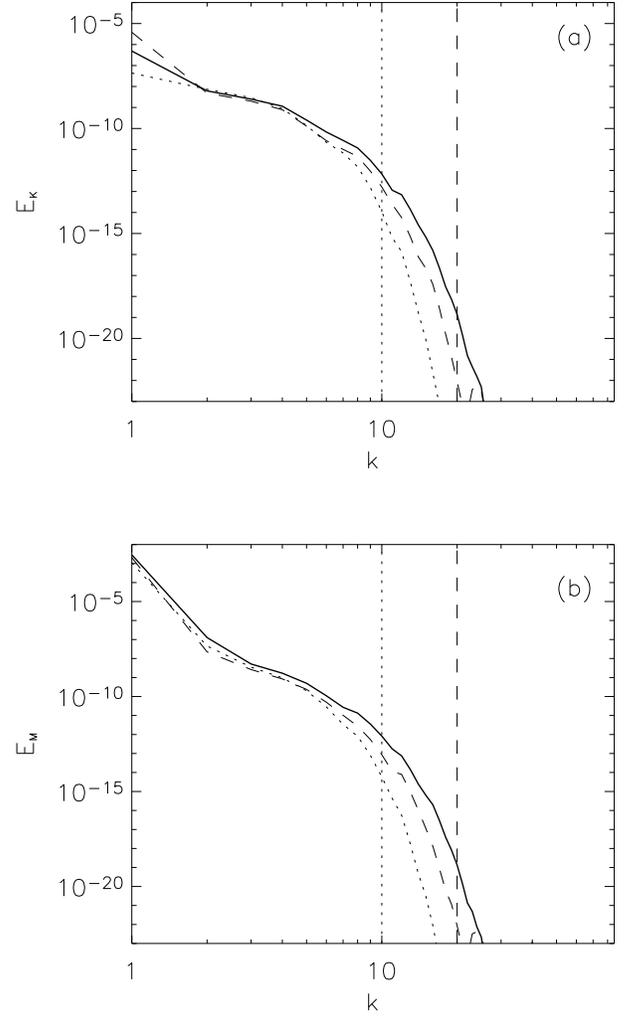}
\caption{(a) Kinetic and (b) magnetic energy spectra, for the three 
    dynamic alignment runs (labels are as in Fig. \ref{fig:decay_E}), 
    at $t=733$.}
\label{fig:decay_latespec}
\end{figure}

Figs. \ref{fig:decay_earlyspec} and \ref{fig:decay_latespec} show the 
associated energy spectra plotted vs. wave number.  Fig. 
\ref{fig:decay_earlyspec} is at an early time ($t=10$) and shows 
the total energy spectrum compensated by Kolmogorov's $-5/3$ law. 
Fig. \ref{fig:decay_latespec} shows kinetic ($E_K$) and magnetic ($E_M$) 
energy spectra at a very late time ($t=733$). The two values of 
$\alpha^{-1}$ are shown as vertical lines. Below $k \sim \alpha^{-1}$, 
the DNS and $\alpha$-model agree reasonably well.

\begin{figure}
\includegraphics[width=8cm]{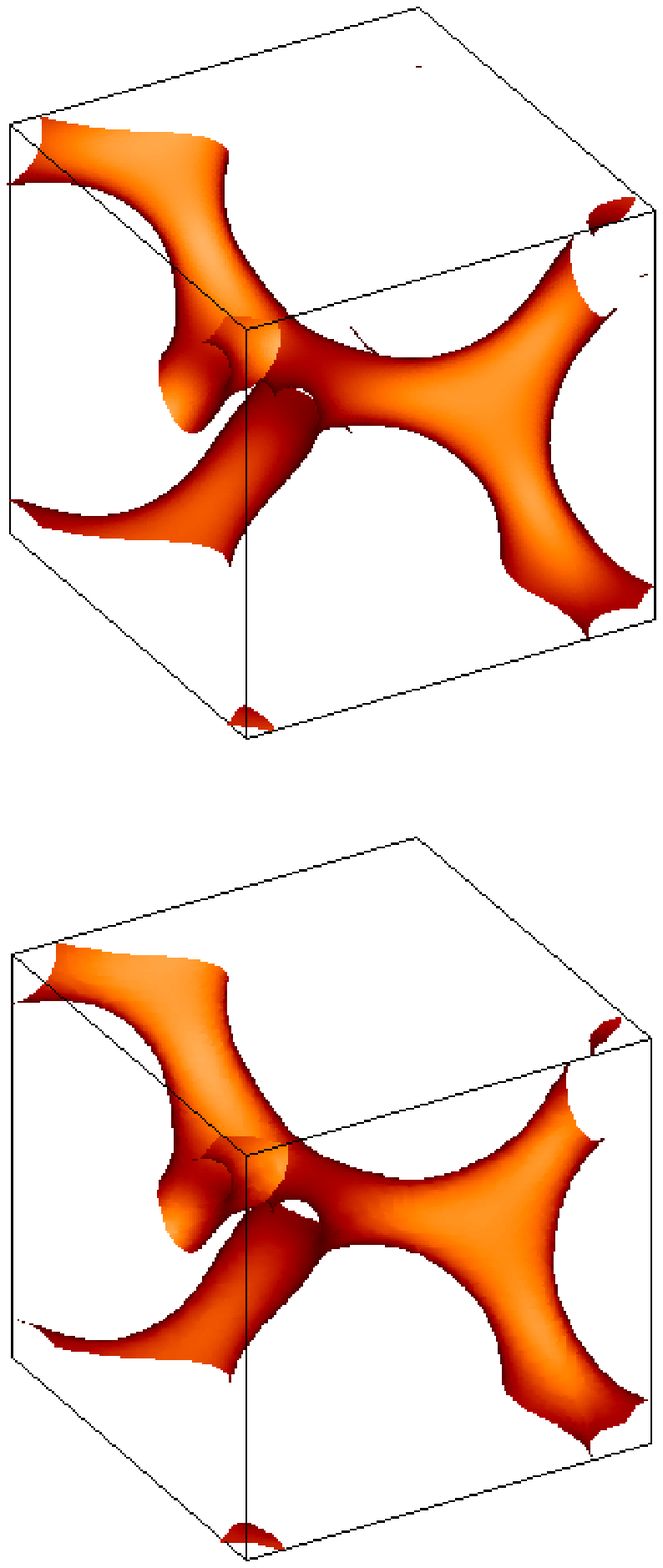}
\caption{(color online) Surfaces of constant magnetic helicity 
    density at $t=800$ at 90\% of its maximum value, for the DNS 
    (above), and the $64^3$ alpha-model simulation (below).}
\label{fig:decay_3D}
\end{figure}

As follows from Figs. \ref{fig:decay_Elog} and 
\ref{fig:decay_latespec}, at late times the magnetic field is 
concentrated at large scales ($k=1$) and has maximum relative 
helicity (note that $E \sim E_M \sim H_M$ after $t \sim 200$ 
in both the DNS and alpha-model simulations). Fig. 
\ref{fig:decay_3D} shows surfaces of constant $H_M$ at $t=800$ 
in the 3D domain, for the DNS and the $64^3$ alpha-model 
simulation. The alpha-model is able to reproduce the large scale 
structures observed in the DNS, and only slight differences can be 
observed. As will be shown in Section \ref{sec:align} this is not 
always the case when using periodic boundary conditions (similar 
results were obtained in 2D MHD simulations \cite{Mininni04}).

\section{\label{sec:align}DYNAMIC ALIGNMENT}

In this case, we load Fourier coefficients into the spherical shells with 
$k_{bot} = 6 \le k \le k_{top} = 10$ with equal amplitudes but 
enough correlation between the phases of $\vv$ and $\vB$ so that 
initially $\left< \vv \cdot \vB \right> = 0.3 \left< |\vv||\vB| \right>$; 
otherwise the phases are random. We again do a $256^3$ DNS run, an 
$\alpha$-model run at $128^3$ with $\alpha=1/20$, and another 
$\alpha$-model run at $64^3$ with $\alpha=1/10$. For all three runs, 
$\nu = \eta = 0.002$. The same conventions are adopted for the graphics 
as in Sec. \ref{sec:decay}.

\begin{figure}
\includegraphics[width=9cm]{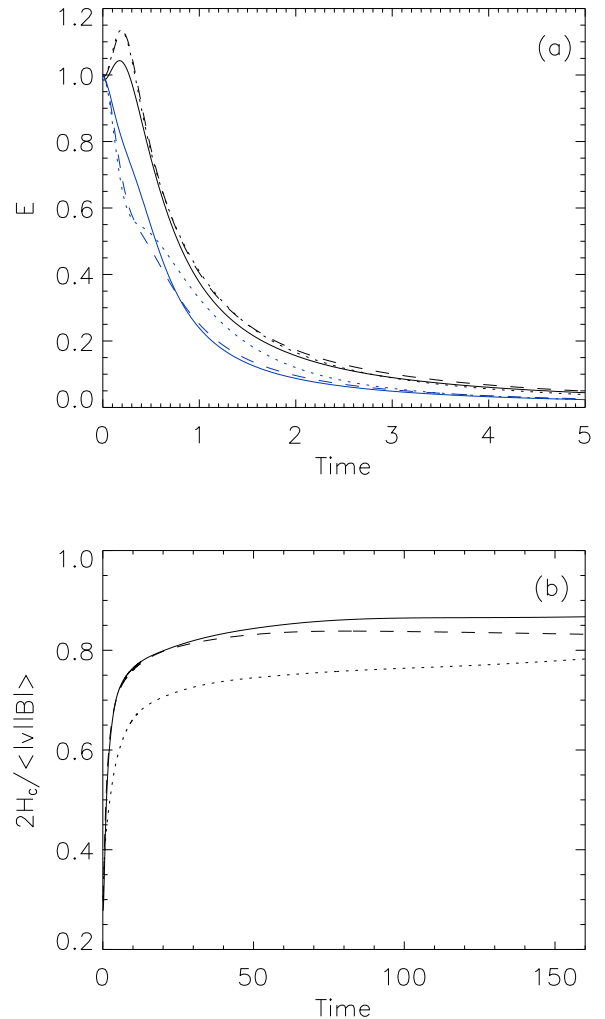}
\caption{(color online) (a) Magnetic energy (upper curves) and kinetic 
     energy (lower blue curves) as a function of time until $t=5$, and 
     (b) relative cross helicity as a function of time until $t=160$, 
     for the dynamic alignment runs. Solid lines correspond to DNS, 
     dashed lines to $128^3$ $\alpha$-model simulations, and dotted 
     lines to $64^3$ $\alpha$-model simulations.}
\label{fig:align_E}
\end{figure}

\begin{figure}
\includegraphics[width=9cm]{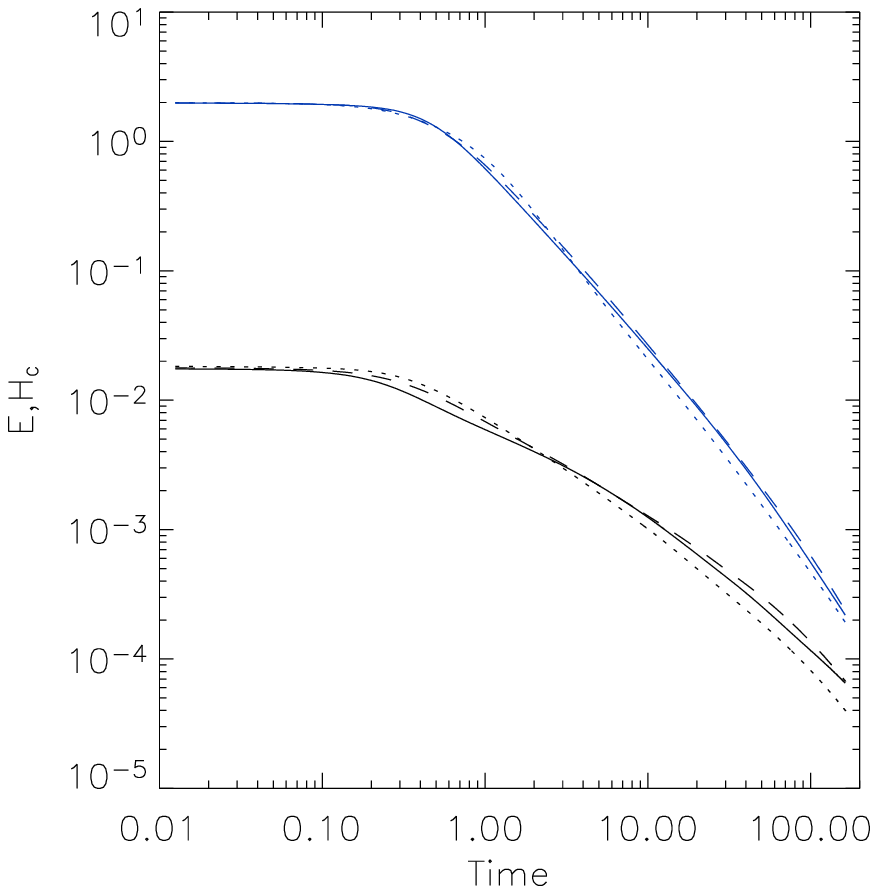}
\caption{(color online) Total energy (upper blue curves) and cross 
     helicity (lower curves) as a function of time. Labels are as in 
     Fig. \ref{fig:align_E}.}
\label{fig:align_Elog}
\end{figure}

Figs. \ref{fig:align_E}a,b show the decay of the kinetic and magnetic 
energies (a), chosen initially to be equal; and (b) the degree of 
alignment, as measured by the mean cosine of the alignment angle, 
$\left< \vu_s \cdot \vB \right>/ \left< |\vu_s||\vB| \right>$ that 
develops as a function of time. Since much of the alignment is 
contributed by the small scales, the $\alpha$-model underestimates 
the degree of alignment, and the disparity becomes greater as 
$\alpha^{-1}$ is decreased, though the accuracy remains within 
the 10 percent level. Fig. \ref{fig:align_Elog} shows the decay 
of both $E$ and $H_c$, with the more rapid decay of the former. 
There is, in this case, no preferential migration of any global 
quantity to long wavelengths.

\begin{figure}
\includegraphics[width=9cm]{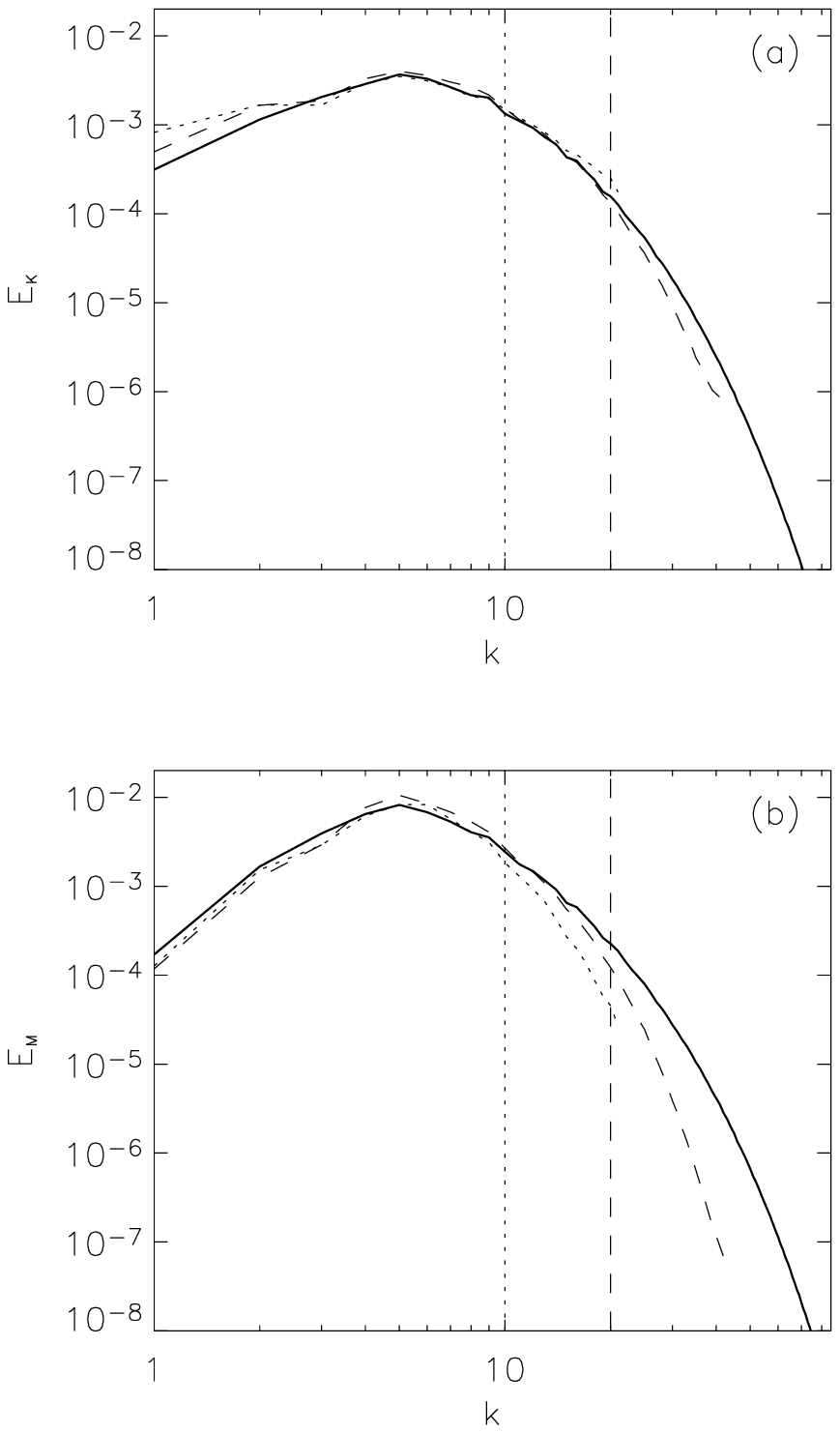}
\caption{(a) Kinetic and (b) magnetic energy spectra, for the three 
    dynamic alignment runs (labels are as in Fig. \ref{fig:align_E}), 
    at $t=4.5$. Vertical dotted and dashed lines indicate respectively 
    the scales $\alpha^{-1} = 10$ and $20$.}
\label{fig:align_earlyspec}
\end{figure}

\begin{figure}
\includegraphics[width=9cm]{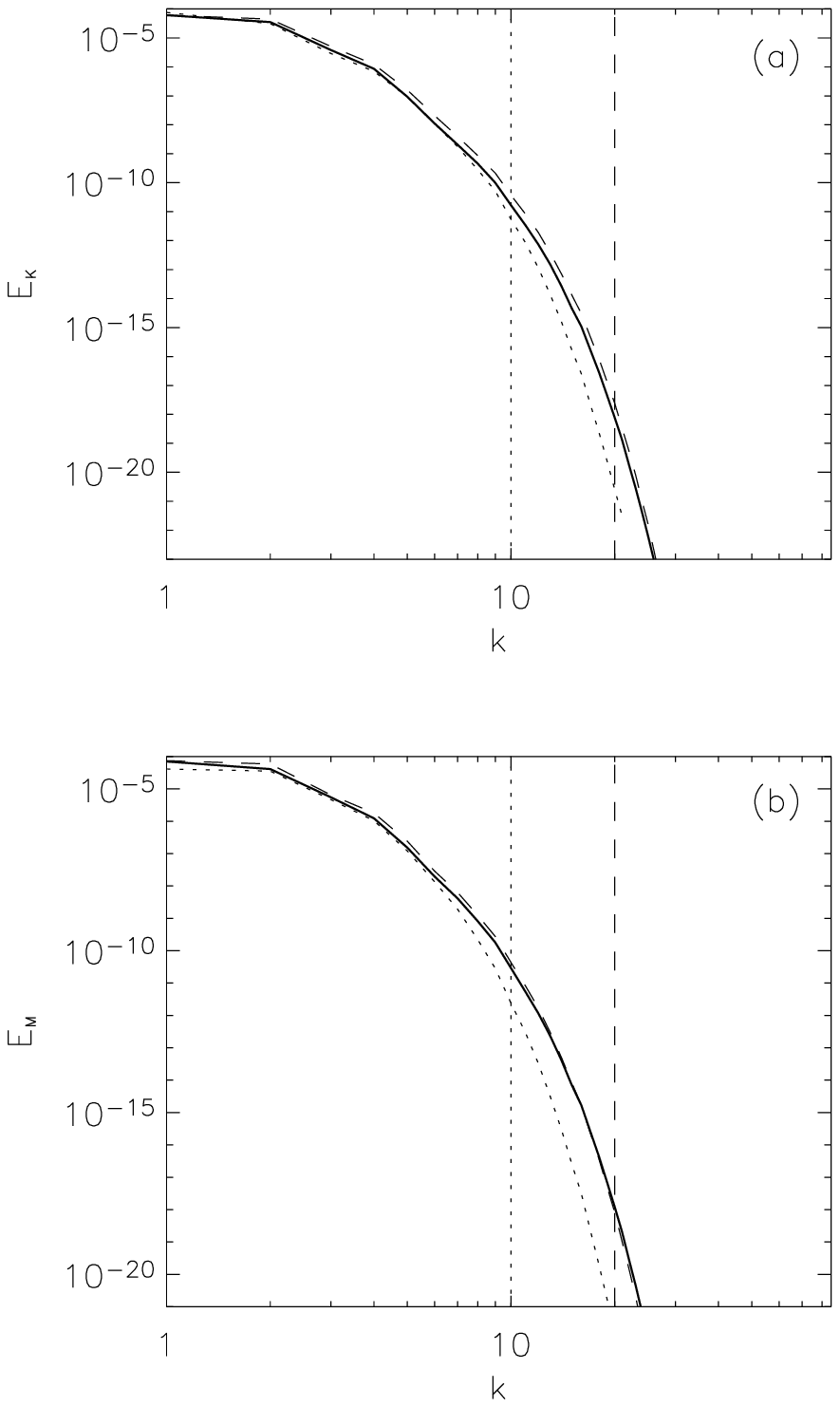}
\caption{(a) Kinetic and (b) magnetic energy spectra, for the three 
    dynamic alignment runs.}
\label{fig:align_latespec}
\end{figure}

Figures \ref{fig:align_earlyspec} and \ref{fig:align_latespec} show 
the kinetic and magnetic energy spectra at an early time, $t=4.5$, 
and at a late one, $t=156$. The agreement of the $\alpha$-model and 
DNS for $k \lesssim \alpha^{-1}$ is again seen to be excellent except 
for an unexplained over-estimate at the earlier time $t=4.5$ for the 
kinetic energy spectrum.

\begin{figure}
\includegraphics[width=8cm]{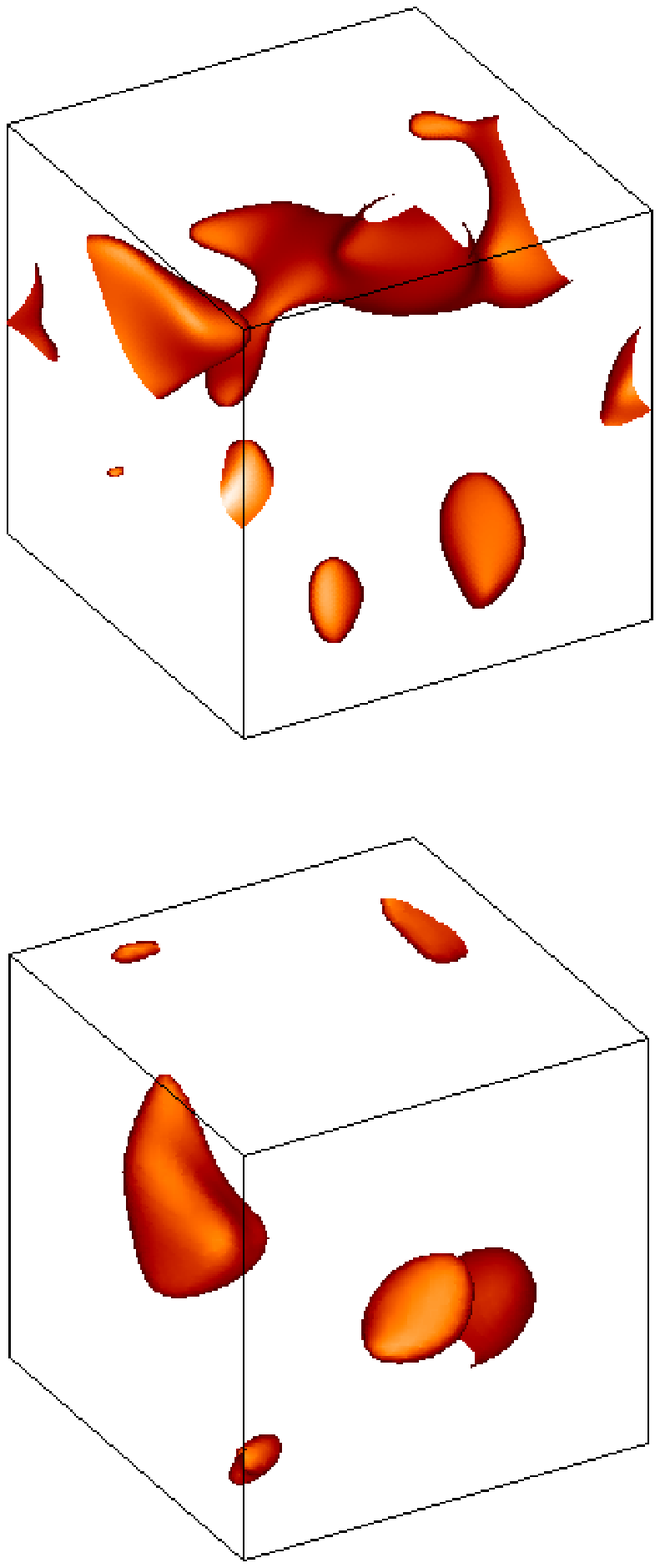}
\caption{(color online) Surfaces of constant cross helicity density at 
    $t=150$ at 50\% of its maximum value, for the DNS (above), and the 
    $64^3$ alpha-model simulation (below).}
\label{fig:align_3D}
\end{figure}

Fig. \ref{fig:align_3D} shows surfaces of constant $H_C$ at 
$t=150$ in the 3D domain, for the DNS and the $64^3$ alpha-model 
simulation. While there are marked similarities in the kinds of 
structures present in the DNS and in the alpha runs, there are
no one-to-one correspondences as to specific features, either 
as to location or orientation. As in the 2D case \cite{Mininni04} we 
conclude that in this case the alpha-model does an excellent job 
reproducing the statistical properties of the large-scale spectra, 
but small-scale detailed phase information (such as the location of 
structures) is lost.

The reason for the striking agreement between the $\alpha$-model and 
DNS exhibited in Fig. \ref{fig:decay_3D}, as contrasted with the 
disagreement shown later in Fig. \ref{fig:align_3D} is that in the case 
of selective decay, both computations have found the same final 
state: the isotropic, maximum-helicity, $k=1$ state. This state 
is the ``ABC flow'' with $A$, $B$, and $C$ all equal.

\section{\label{sec:inverse}INVERSE CASCADES}

Inverse cascades of magnetic helicity, driven mechanically at the 
small scales, have long been known to be an efficient dynamo 
mechanism for generating large-scale magnetic fields \cite{Meneguzzi81}. 
Here, we try  a different approach: we drive the magnetic field directly 
at small scales with a random forcing function that is a superposition 
of ``ABC'' flows [see Eq. (\ref{eq:ABC})] between wavenumbers $k=8$ and 
9, and with no driving for the velocity. The randomness is introduced 
by randomly changing the phases in the trigonometric arguments of each 
ABC component with a correlation time of 
$\Delta t = 1.25 \times 10^{-2}$ (in all the simulations we discuss 
in this section, the time step is $2.5 \times 10^{-3}$). A tiny seed 
velocity field is amplified somewhat, but the kinetic energy always 
remains well below the level of the magnetic energy throughout. We 
again exhibit the results of a $256^3$ DNS computation, and 
$\alpha$-model computations with $\alpha=1/20$ and $1/10$, with 
resolutions of $128^3$ and $64^3$ respectively.

\begin{figure}
\includegraphics[width=9cm]{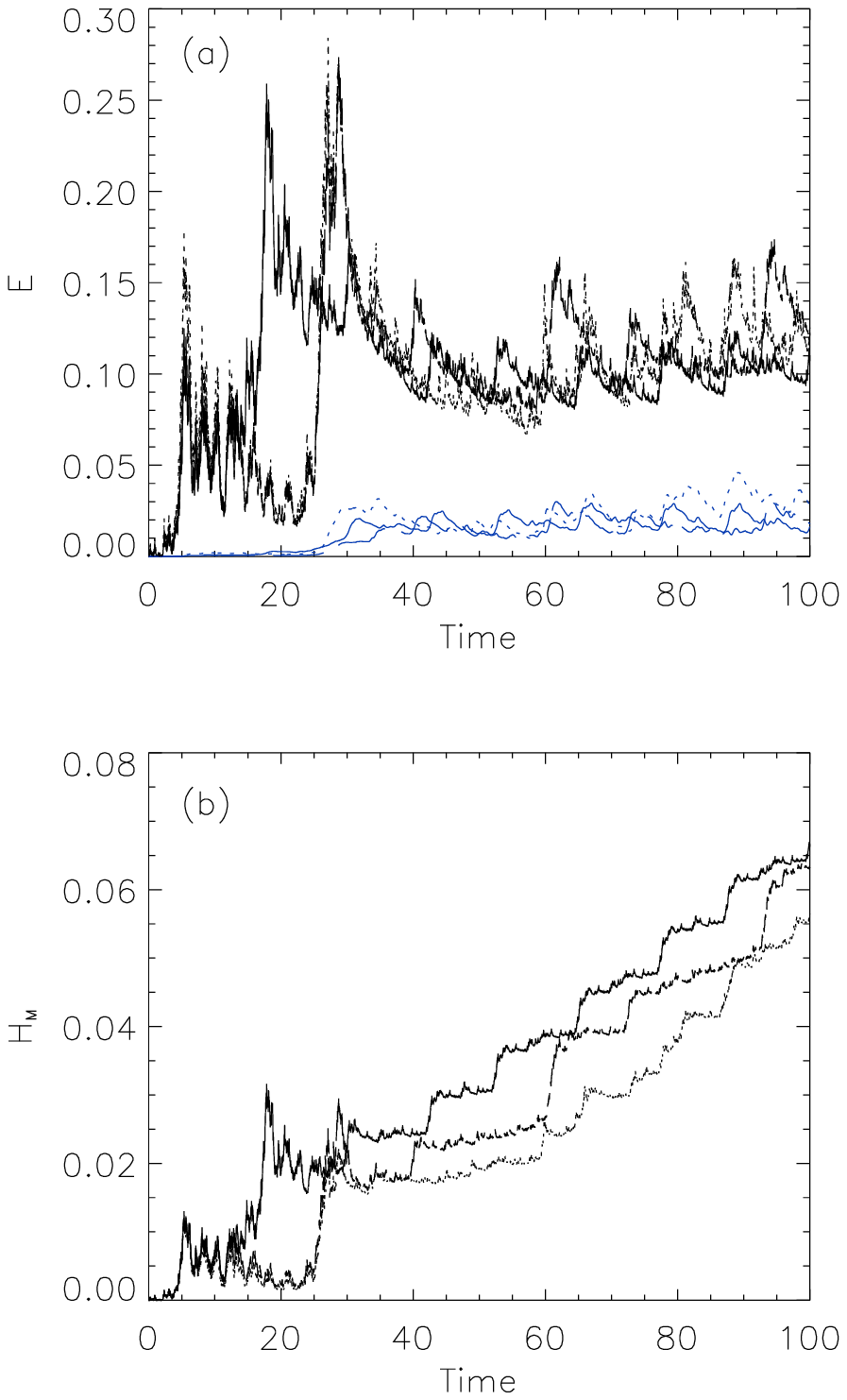}
\caption{(color online) (a) Magnetic energy (upper curves) and kinetic 
     energy (lower blue curves) as a function of time, and (b) magnetic 
     helicity as a function of time. Solid lines correspond to DNS, 
     dashed lines to $128^3$ $\alpha$-model simulations, and dotted 
     lines to $64^3$ $\alpha$-model simulations.}
\label{fig:inver_E}
\end{figure}

\begin{figure}
\includegraphics[width=9cm]{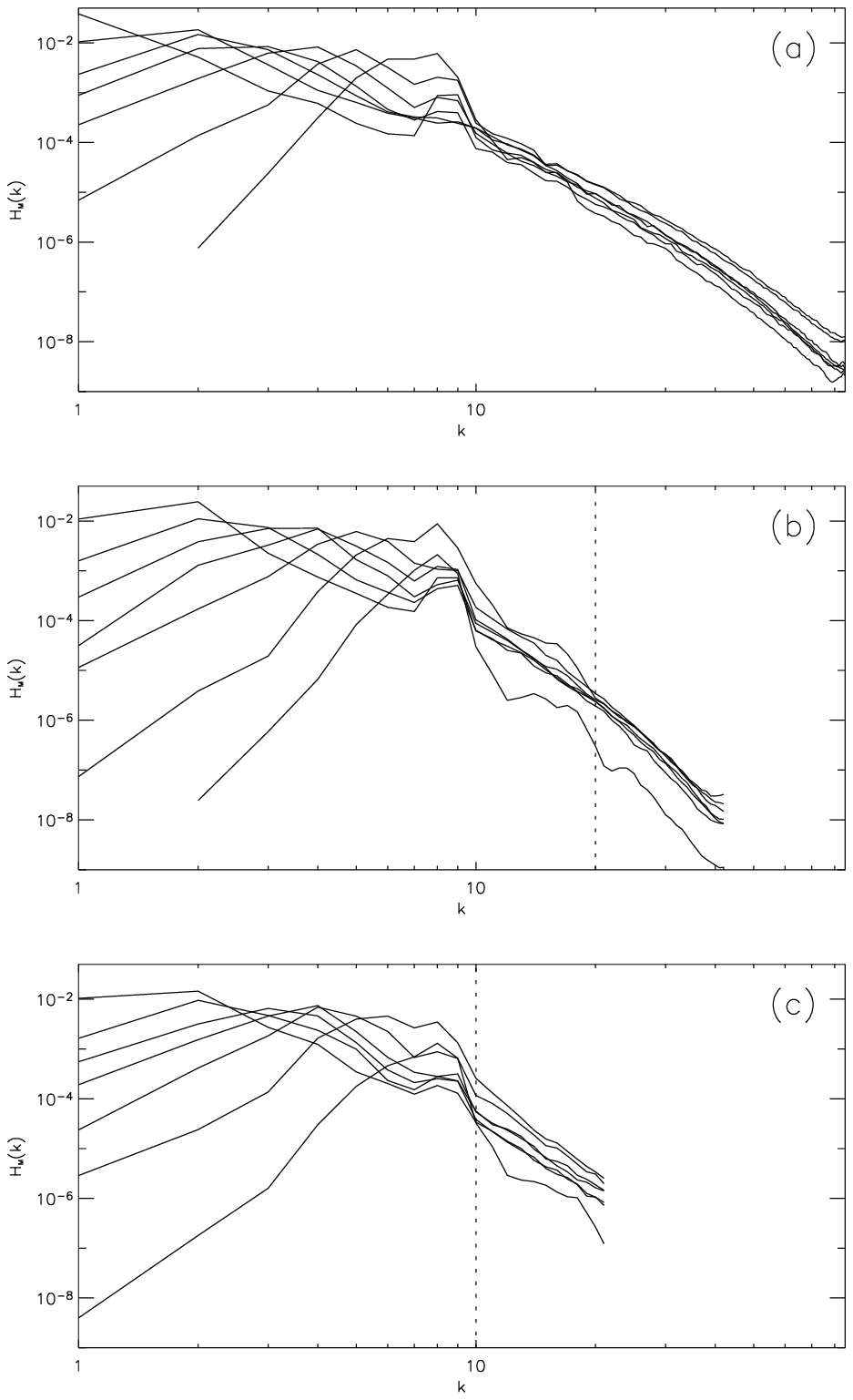}
\caption{Spectrum of magnetic helicity for different times, for 
     $t= 30$, 35, 40, 45, 55, and $72.5$; (a) DNS, (b) $128^3$ 
     $\alpha$-model, and (c) $64^3$ $\alpha$-model. The vertical 
     lines indicate $\alpha^{-1}$. Note the cascade of magnetic 
     helicity to large scales as time evolves.}
\label{fig:inver_timespec}
\end{figure}

Figs. \ref{fig:inver_E} show the time histories of the energies (a) 
and magnetic helicities (b) for the three runs. The rather abrupt 
phase jumps in the ABC flows give the lines a jagged appearance and 
it is sometimes difficult to identify which of the three runs is which. 
Suffice it to say that the two $\alpha$-model runs exhibit the same 
features as the DNS runs, but with a time lag that is greater for the 
larger $\alpha$. This is visible more clearly in Fig. 
\ref{fig:inver_timespec}, where the magnetic helicity spectra for the 
three simulations are plotted as functions of $k$. The curves are 
the helicity spectra as functions of time. The lower levels of 
excitation are associated with earlier times. The times exhibited 
range from $t = 30$ to $t = 72.5$. The peak, once established, moves to 
the left with nearly the same speed in each case. The suppression 
of small scales, where the unsmoothed ABC flow is also unstable, may 
be responsible for the time lag. This time-lagged behavior is 
reminiscent of what happened in two dimensions with the inverse 
cascade of mean square vector potential \cite{Mininni04}. However, 
note that in three dimensions once the inverse cascade has been 
established, the growth rate of magnetic helicity is well captured 
by the alpha-model (Fig. \ref{fig:inver_E}.b), indicative of a more 
local cascade (in scale). The power laws present in the spectra 
of magnetic helicity, and kinetic and magnetic energy \cite{Pouquet76} 
are also well captured by the alpha-model.

\section{\label{sec:dynamo}THE DYNAMO}

The mechanically-driven dynamo, in which injected mechanical 
energy is converted to magnetic energy at large scales, has long 
been a recurrent problem in MHD \cite{Brandenburg04}. Here we are 
able to show that the alpha model yields the same results within 
acceptable accuracy as those of a DNS of the same situation (see 
Ref. \cite{Ponty04} for another case of recent interest).

We begin with a velocity field which is again forced externally 
with the ``ABC'' geometry of Eq. (\ref{eq:ABC}). We choose $A=0.9$, 
$B=1.0$, and $C=1.1$, $k_0=3$, with $\eta=\nu=0.002$. This choice 
is governed by the knowledge that the $A=B=C$ flow gives the largest 
dynamo growth rate \cite{Galanti92} but it is hydrodynamically very 
stable \cite{Podvigina94}; breaking that symmetry allows for 
turbulence to develop faster \cite{Archontis03}.

The force is allowed to operate until a statistically-steady turbulent 
Navier-Stokes flow prevails. Then a magnetic seed field is 
introduced at a very low level in the modes from $k=1$ to $k=10$. 
As in some other sections, we compare a DNS run at resolution 
$256^3$ with two $\alpha$-model runs, one with $\alpha=1/20$ and 
$128^3$ resolution, the other with $\alpha=1/10$ and $64^3$ resolution. 

Before embarking on the MHD comparison between DNS and alpha-model 
results, it is instructive to compare the hydrodynamic properties 
of the flow. When the small magnetic seed is introduced, the 
Lorentz force in the Navier-Stokes equation can be neglected. 
The induction equation is linear in the magnetic field, and as a 
result, the geometrical properties of the flow are responsible for 
the observed amplification.

\begin{figure}
\includegraphics[width=9cm]{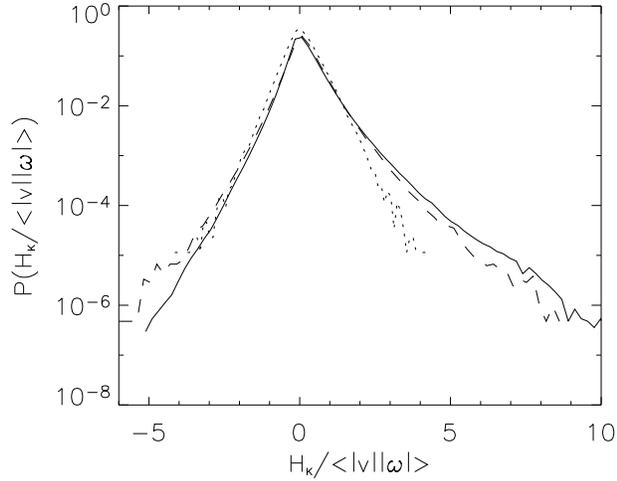}
\caption{Probability distribution function of relative kinetic 
     helicity.  Solid lines correspond to DNS, dashed lines to 
     $128^3$ $\alpha$-model simulations, and dotted lines to $64^3$ 
     $\alpha$-model simulations.}
\label{fig:dynamo_Hpdf}
\end{figure}

The flow generated by the external ABC force is helical. Previous 
studies of the alpha-model behavior in simulations of hydrodynamic 
flows were carried for non-helical flows \cite{Chen99c,Mohseni03}. 
As a result, here we will focus only on the characterization of the 
flow helicity. The amount of helicity in a flow (both for DNS and the 
alpha-model \cite{Foias01}) is measured by the kinetic helicity 
\begin{equation}
H_K = \frac{1}{2} \int \vv \cdot \vomega \, \textrm{d}^3x \; .
\end{equation}
It is also useful to normalize this quantity introducing the relative 
helicity $2 H_K /(\left<|v|\right> \left<|\omega|\right>)$. 
Fig. \ref{fig:dynamo_Hpdf} shows the probability distribution function 
(pdf) of relative kinetic helicity for the DNS and alpha-model 
simulations. A stronger positive tail can be identified in all cases, 
giving rise to a net positive kinetic helicity in the flow.

\begin{figure}
\includegraphics[width=9cm]{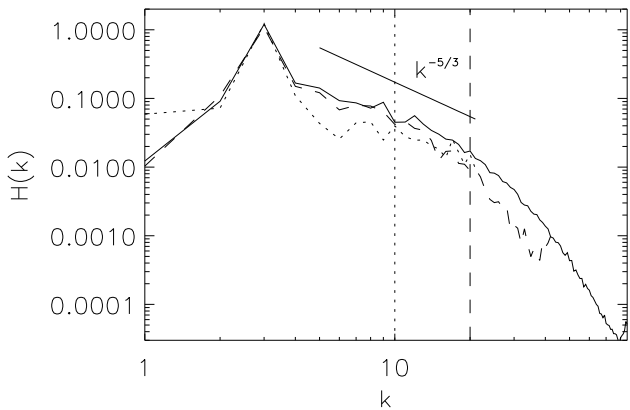}
\caption{Spectrum of kinetic helicity. The Kolmogorov's slope is 
     shown as a reference. The vertical lines indicate $\alpha^{-1}$. 
     Labels are as in Fig. \ref{fig:dynamo_Hpdf}.}
\label{fig:dynamo_Hspec}
\end{figure}

In 3D hydrodynamic turbulence, kinetic helicity is an ideal invariant 
and is known to cascade to smaller scales \cite{Brissaud73,Andre77}. 
Fig. \ref{fig:dynamo_Hspec} shows the spectrum of $H_K$ during the 
hydrodynamic simulation. As with the energy, the alpha-model is able 
to capture the evolution of kinetic helicity in Fourier space up to 
$k \sim \alpha^{-1}$. It seems that a Kolmogorov spectrum results for 
helicity \cite{Chen03,Gomez04}, which implies that the relative helicity 
is weaker at small scales.

\begin{figure}
\includegraphics[width=9cm]{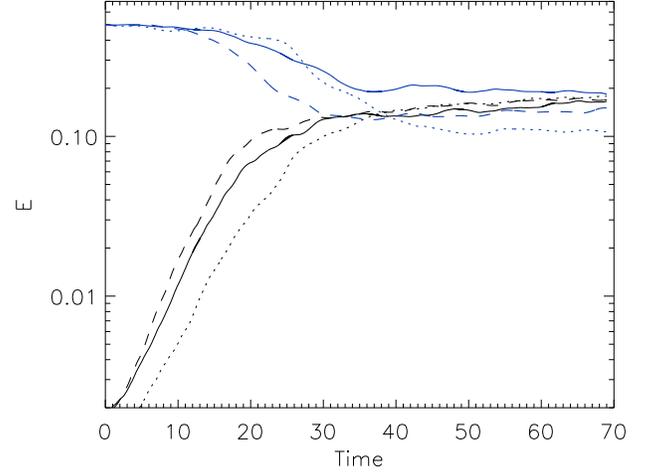}
\caption{(color online) Kinetic energy (upper blue curves) and magnetic 
     energy (lower curves) as a function of time. Labels are as in 
     Fig. \ref{fig:dynamo_Hpdf}.}
\label{fig:dynamo_E}
\end{figure}

The early stages of the growth of the magnetic field are in the 
``kinematic dynamo'' parameter regime, involving exponential growth 
of the magnetic energy. This is shown in Fig. \ref{fig:dynamo_E}, 
which exhibits both the kinetic and magnetic energy as functions 
of time for the three runs. Though the three energies as functions 
of time are offset, it is clear that the linear growth rates are 
close. At about $t=30$, there is a saturation, close to a state 
in which on the average the energy is equipartitioned approximately 
between kinetic and magnetic. After that, there are no significant 
variations in the evolution of the total kinetic and magnetic 
energy.

\begin{figure}
\includegraphics[width=9cm]{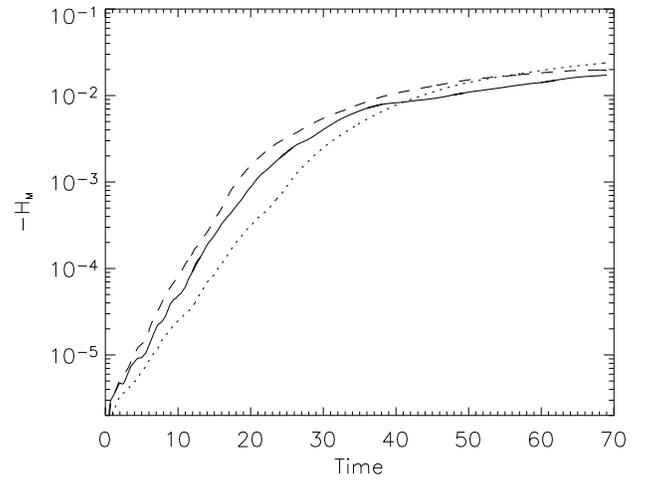}
\caption{Negative of the magnetic helicity as a function of time. 
     Labels are as in Fig. \ref{fig:dynamo_Hpdf}.}
\label{fig:dynamo_mhel}
\end{figure}

\begin{figure}
\includegraphics[width=9cm]{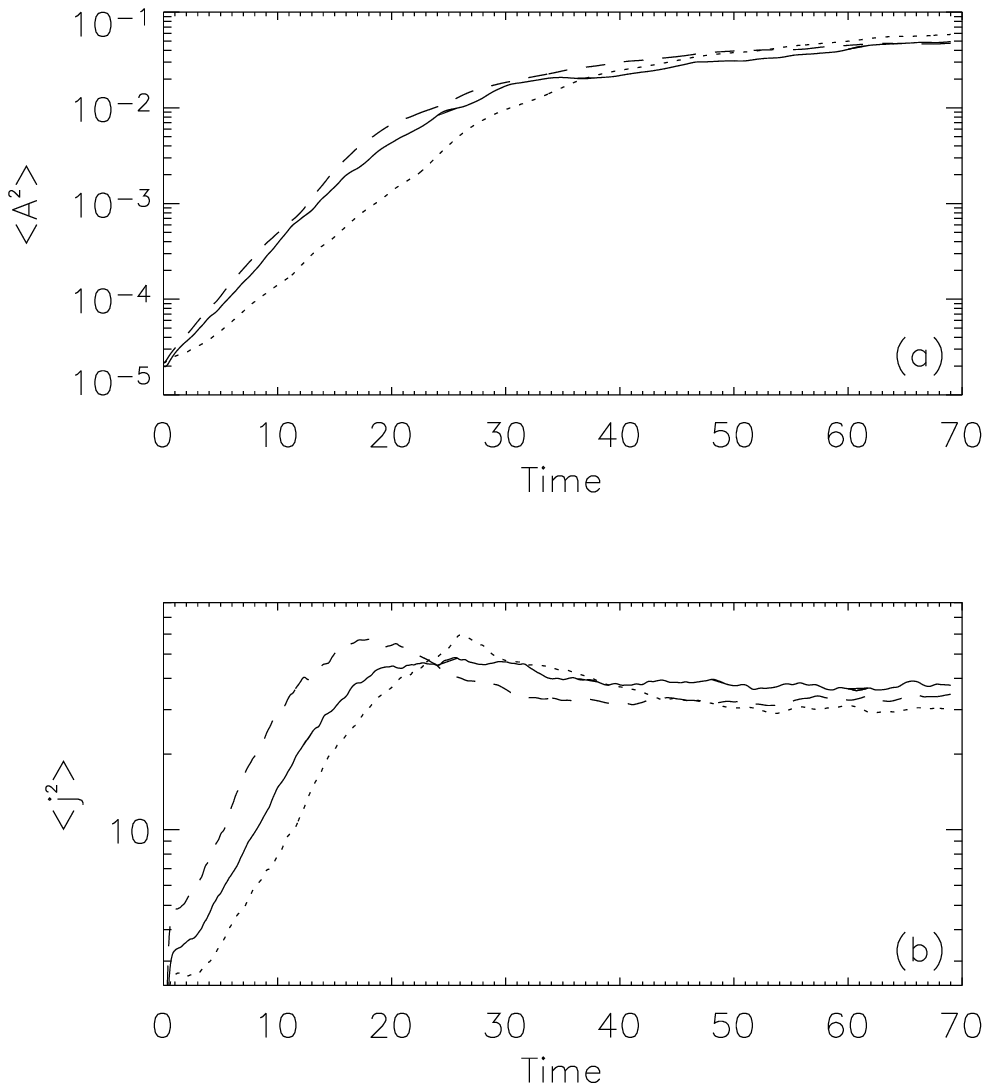}
\caption{(a) Mean square vector potential, and (b) mean square 
     current density as functions of time. Labels are as in Fig. 
     \ref{fig:dynamo_Hpdf}.}
\label{fig:dynamo_curr}
\end{figure}

Figs. \ref{fig:dynamo_mhel} and \ref{fig:dynamo_curr}.a,b show the 
negative of the magnetic helicity, the mean square vector potential, 
and the mean square current density as functions of time. Though the 
agreements are not sharp, it is clear that the saturation levels and 
the times of saturation are both well approximated. Note that, in 
accord with expectations \cite{Brandenburg01}, the magnetic helicity 
acquires a negative value, opposite to the sign of the injected 
mechanical helicity. Note also that growth rates of both small scales 
(represented by the square current) and large scales (represented by 
the square vector potential) are well approximated by the alpha model 
during the kinematic regime.

While at $t \approx 30$ saturation in the exponential growth of 
magnetic energy takes place, the large scale modes continue growing, 
and at the end the magnetic field is dominated by large scales. While 
the mean square current density is constant after $t=30$, the 
squared vector potential keeps growing slowly. This behavior is even 
clearer in the evolution of the magnetic energy spectrum.

\begin{figure}
\includegraphics[width=9cm]{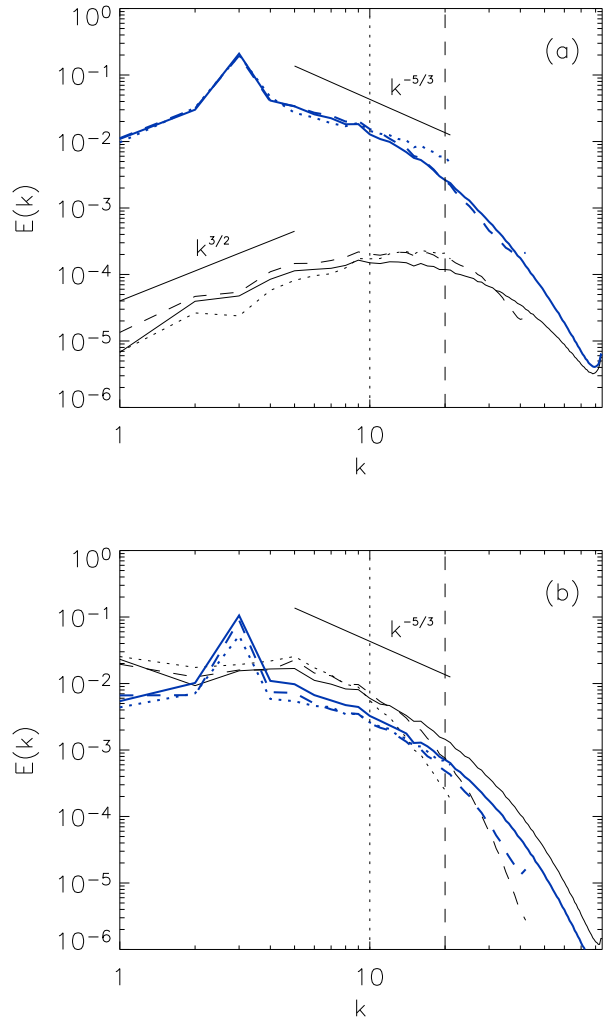}
\caption{(color online) Kinetic (thick blue lines), and magnetic energy 
    spectra (thin lines), at (a) $t=6$ and (b) $t=70$. Kolmogorov's 
    $k^{-5/3}$ and Kazantsev's $k^{3/2}$ spectra are shown as a 
    reference. The vertical dotted and dashed lines correspond to the 
    scales $\alpha^{-1} = 10$ and 20 respectively. Labels are as in 
    Fig. \ref{fig:dynamo_Hpdf}.}
\label{fig:dynamo_spec}
\end{figure}

Figs. \ref{fig:dynamo_spec}.a,b show the evolution of the kinetic 
and magnetic spectra. The thick lines indicate kinetic spectra and 
the thin lines the magnetic spectra. The vertical lines indicate the 
locations of the two values of $\alpha^{-1}$. In Fig. 
\ref{fig:dynamo_spec}.a, the upper traces are the DNS spectrum at the 
time the seed field begins to grow, both for DNS and $\alpha$-model 
simulations. The lower traces in Fig. \ref{fig:dynamo_spec}.a show the 
magnetic spectrum at an early stage of its evolution. During this 
stage, the magnetic energy spectrum peaks at small scales, and the 
$\alpha$-model correctly captures the overall shape of the spectrum as 
well as the scale where the magnetic energy peaks. In the kinematic 
regime, all the magnetic $k$-shells in Fourier space (up to 
$k \lesssim 12$) grow with the same rate, and this feature is also 
well captured by the $\alpha$-model simulations (not shown). This 
evolution is characteristic of small scale dynamos, as well as a 
$k^{3/2}$ slope in the magnetic energy spectrum at early times 
\cite{Kazantsev67,Brandenburg01}. Fig. \ref{fig:dynamo_spec}.b shows 
the late-time spectra, when approximate equipartition has been achieved. 
Note that as a result of helical dynamo action, a magnetic field at 
large scales ($k=1$) is generated (see Fig. \ref{fig:dynamo_spec}.b). 
The amplitude of this mode is in good agreement for both DNS and 
$\alpha$-model simulations.

\begin{figure}
\includegraphics[width=8cm]{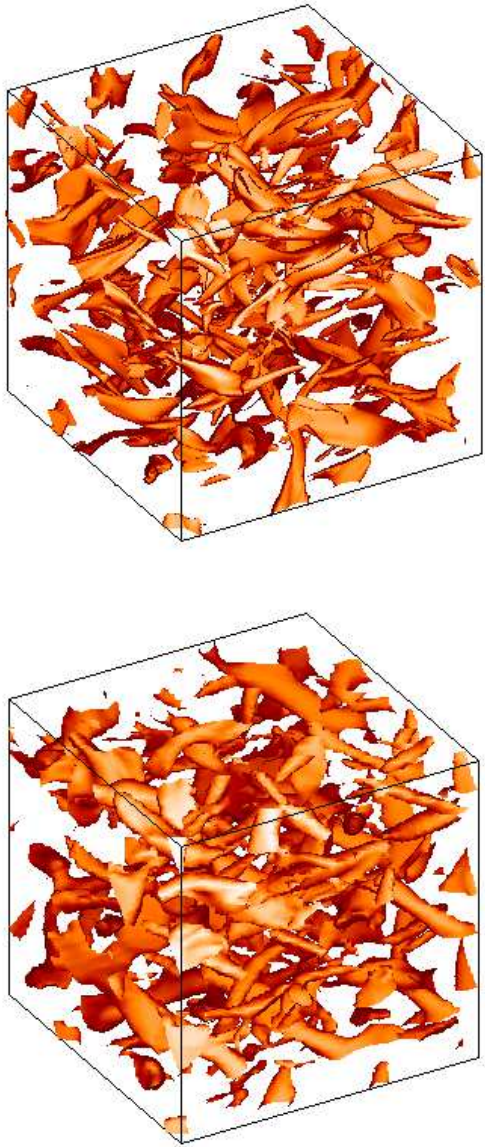}
\caption{(color online) Surfaces of constant magnetic energy at 
    $t=60$ at 50\% of its maximum value, for the DNS (above), and the 
    $64^3$ alpha-model (below).}
\label{fig:dynamo_3D}
\end{figure}

Figure \ref{fig:dynamo_3D} shows surfaces of constant magnetic energy 
at $t=60$, when the nonlinear saturation has already taken place but the 
large scale magnetic field is still growing. Thin and elongated structures 
can be identified in the magnetic field growing in the DNS. However, 
note that while these structures are present both in the DNS and in the 
alpha-model, in the latter case the structures are thicker. This change 
is related to the filtering length $\alpha$ in the alpha-model. Similar 
results have been found in vorticity structures observed in simulations 
of hydrodynamic turbulence using the alpha-model \cite{Chen99c}.

We thus conclude that there are few surprises in the dynamo simulations, 
at least for these values of $\eta/\nu$, and no glaring departures of 
the $\alpha$-model predictions from the DNS results.

\section{\label{sec:sum}SUMMARY; DISCUSSION}

Within the framework of rectangular periodic boundary conditions, we 
have examined four familiar three-dimensional MHD turbulence effects 
via the $\alpha$-model and DNS. In every case, the principal large-scale 
features of these phenomena have been achieved with the $\alpha$-model 
to acceptable accuracy. The savings in computer time achieved by the 
$\alpha$-model runs have ranged from $256^3/128^3 = 8$ to 
$256^3/64^3 = 64$, without considering extra saving in the time step 
from the CFL condition as the resolution is decreased. In no case has 
the $\alpha$-model yielded results at significant variance with the 
DNS runs, which have been regarded as accurate.

Other features of the DNS runs, such as the probability distribution 
functions of the fluctuating quantities (such as local energy 
dissipation rates), have also been reproduced by the $\alpha$-model 
as they were in two dimensions \cite{Mininni04}, but we have not 
shown those results here because they are so similar to what 
was found in two dimensions.

In Ref. \cite{Mininni04} also the errors of the $\alpha$-model 
computations were compared against under-resolved DNS. The behavior 
of the $\alpha$-model in three dimensions is comparable to our previous 
results, and therefore we refer the reader to our previous work for a 
detailed discussion about this topic.

In conclusion, the MHD $\alpha$-model can be considered to be validated, 
at least for the behavior of long-wavelength spectra in periodic boundary 
conditions. Its implementation in the presence of material boundaries 
stands as a next forbidding challenge.

\begin{acknowledgments}
We thank H. Tufo for providing computer time at UC-Boulder, NSF ARI 
grant CDA-9601817. Computer time was also provided by NCAR and 
Dartmouth. The NSF grants ATM-0327533 at Dartmouth College and 
CMG-0327888 at NCAR supported this work in part and are gratefully 
acknowledged.
\end{acknowledgments}



\begin{thebibliography}{1}

\bibitem{Holm98a}
D.D. Holm, J.E. Marsden and T.S. Ratiu,
``The Euler-Poincar\'e Equations and Semidirect Products with 
Applications to Continuum Theories,''
{\it Adv. in Math.} {\bf 137}, 1-81 (1998).

\bibitem{Holm98b}
D.D. Holm, J.E. Marsden and T.S. Ratiu,
``Euler-Poincar\'e Models of Ideal Fluids with Nonlinear Dispersion,'' 
{\it Phys. Rev. Lett.} {\bf 80}, 4173-4176 (1998).

\bibitem{Chen98}
S.Y. Chen, D.D. Holm, C. Foias, E.J. Olson, E.S. Titi, and S. Wynne,
``The Camassa-Holm equations as a closure model for turbulent channel 
and pipe flows,'' 
{\it Phys. Rev. Lett.} {\bf 81}, 5338-5341 (1998).

\bibitem{Chen99a}
S.Y. Chen, C. Foias, D.D. Holm, E. Olson, E.S. Titi, S. Wynne,
``The Camassa-Holm equations and turbulence,''
{\it Physica D} {\bf 133} 49-65 (1999).

\bibitem{Chen99b}
S.Y. Chen, C. Foias, D.D. Holm, E.J. Olson, E.S. Titi, and S. Wynne, 
``A connection between the Camassa-Holm equations and turbulence in 
pipes and channels,''
{\it Phys.  Fluids}  {\bf 11}, 2343-2353 (1999).

\bibitem{Chen99c}
S.Y. Chen, D.D. Holm, L.G. Margolin, and R. Zhang,
``Direct numerical simulations of the Navier-Stokes alpha model,''
{\it Physica D} {\bf 133}, 66-83 (1999).

\bibitem{Foias01}
C. Foias, D.D. Holm, and E.S. Titi,
``The Navier-Stokes-alpha model of fluid turbulence,'' 
{\it Physica D} {\bf 152}, 505-519 (2001). 

\bibitem{Nadiga01}
B. T. Nadiga and S. Shkoller, 
``Enhancement of the inverse cascade of energy in the two-dimensional
Lagrangian-averaged Navier-Stokes equations,''
{\it Phys. Fluids} {\bf 13}, 1528-1531 (2001).

\bibitem{Holm02}
D.D. Holm, 
``Averaged Lagrangians and the mean dynamical effects of 
fluctuations in ideal fluid dynamics,''
{\it Physica D} {\bf 170}, 253-286 (2002).

\bibitem{Holm02b}
D. D. Holm,
``Lagrangian averages, averaged Lagrangians, and the mean effects of
fluctuations in fluid dynamics,''
{\it Chaos} {\bf 12}, 518-530 (2002).

\bibitem{Foias02}
C. Foias, D.D. Holm, and E.S. Titi, 
``The three-dimensional viscous Camassa-Holm equations and their 
relation to the Navier-Stokes equations and turbulence theory,''
{\it J. Dynamics and Diff. Equ.} {\bf 14}, 1-35 (2002).

\bibitem{Ilyin03}
A.A. Ilyin and E.S. Titi, 
``Attractors to the two-dimensional Navier-Stokes-$\alpha$ model: 
An $\alpha$-dependence study,'' 
{\it J. Dynamics Diff. Equ.} {\bf 15}, 751-777 (2003).

\bibitem{Mohseni03}
K. Mohseni, B. Kosovi\'c, S. Shkoller, and J.E. Marsden, 
``Numerical simulations of the Lagrangian averaged Navier-Stokes 
equations for homogeneous isotropic turbulence,'' 
{\it Phys. Fluids} {\bf 15}, 524-544 (2003).

\bibitem{Mininni04}
P.D. Mininni, D.C. Montgomery, and A.G. Pouquet,
``A numerical study of the alpha model for two-dimensional 
magnetohydrodynamic turbulent flows,''
{\it Phys. Fluids}, submitted (arXiv:physics/0410159).

\bibitem{Ponty04}
Y. Ponty, P.D. Mininni, D.C. Montgomery, J.-F. Pinton, H. Politano, and 
A. Pouquet,
``Numerical study of dynamo action at low magnetic Prandtl numbers,''
{\it Phys. Rev. Lett.}, submitted (arXiv:physics/0410046).

\bibitem{Montgomery02}
D. Montgomery and A. Pouquet, 
``An alternative interpretation for the Holm `alpha model',''
{\it Phys. Fluids} {\bf 14}, 3365--3366 (2002).

\bibitem{Matthaeus80}
W.H. Matthaeus and D. Montgomery,
``Selective decay hypothesis at high mechanical and magnetic Reynolds 
numbers,''
{\it Ann. N.Y. Acad. Sci.} {\bf 357}, 203 (1980).

\bibitem{Ting86}
A.C. Ting, W.H. Matthaeus, and D. Montgomery,
``Turbulent relaxation processes in magnetohydrodynamics,''
{\it Phys. Fluids} {\bf 29}, 3261 (1986).

\bibitem{Kinney95}
R. Kinney, J.C. McWilliams and T. Tajima,
``Coherent structures and turbulent cascades in two-dimensional
incompressible magnetohydrodynamic turbulence,''
{\it Phys. Fluids} {\bf 2}, 3623-3639 (1995).

\bibitem{Grappin83}
R. Grappin, A. Pouquet, and J. L\'eorat, 
`` Dependence on Correlation of MHD Turbulence Spectra,'' 
{\it Astron. Astrophys.} {\bf 126}, 51-56 (1983).

\bibitem{Pouquet86}
A. Pouquet, M. Meneguzzi, and U. Frisch, 
`` The Growth of Correlations in MHD Turbulence,'' 
{\it Phys. Rev. A} {\bf 33}, 4266-4276 (1986).

\bibitem{Ghosh88}
S. Ghosh, W.H. Matthaeus, and D.C. Montgomery, 
``The evolution of cross helicity in driven/dissipative two-dimensional 
magnetohydrodynamics,'' 
{\it Phys. Fluids} {\bf 31}, 2171-2184 (1988).

\bibitem{Lilly69}
D.K. Lilly, ``Numerical simulation of two-dimensional turbulence,''
{\it Phys. Fluids Suppl. II} {\bf 12}, 240-249 (1969).

\bibitem{Mazure75}
U. Frisch, A. Pouquet, J. L\'eorat, and A. Mazure,
``On the possibility of an inverse cascade in MHD helical turbulence,''
{\it J. Fluid Mech.} {\bf 68}, 769--778 (1975).

\bibitem{Pouquet76}
A. Pouquet, U. Frisch, and J. L\'eorat, 
``Strong MHD helical turbulence and the nonlinear dynamo effect,'' 
{\it J. Fluid. Mech.} {\bf 77}, 321-354 (1976).

\bibitem{Hossain83}
M. Hossain, W.H. Matthaeus, and D. Montgomery, 
``Long-time states of cascades in the presence of a maximum length scale,''
{\it J.Plasma Phys.} {\bf 30}, 479-493 (1983).

\bibitem{Brandenburg04}
A. Brandenburg and K. Subramanian, 
``Astrophysical magnetic fields and nonlinear dynamo theory,'' 
astro-ph/0405052.

\bibitem{Meneguzzi81}
M. Meneguzzi, U. Frisch, and A. Pouquet, 
``Helical and non-helical turbulent dynamos,''
{\it Phys. Rev. Lett.} {\bf 47}, 1060--1064 (1981).

\bibitem{Galanti92}
B. Galanti, P.L. Sulem, and A. Pouquet,
``Linear and non-linear dynamos associated with ABC flows,'' 
{\it Geophys. Astrophys. Fluid Dyn.} {\bf 66}, 183-208 (1992).

\bibitem{Podvigina94} 
O. Podvigina and A. Pouquet, 
``On the non-linear stability of the 1:1:1 ABC flow,'' 
{\it Physica D} {\bf 75}, 471-508 (1994).

\bibitem{Archontis03}
V. Archontis, S.B.F. Dorch, and {\AA}. Nordlund, 
``Dynamo action in turbulent flows,'' 
{\it Astron. Astrophys.} {\bf 410}, 759-766 (2003).

\bibitem{Brissaud73}
A. Brissaud, U. Frisch, J. L\'eorat, M. Lesieur, and A. Mazure, 
``Helicity cascades in fully developed isotropic turbulence,'' 
{\it Phys. Fluids} {\bf 16}, 1366-1367 (1973). 

\bibitem{Andre77}
J.C. Andr\'e and M. Lesieur, 
``Influence of helicity on the evolution of isotropic turbulence at 
high Reynolds number,'' 
{\it J. Fluid Mech.} {\bf 81}, 187-207 (1977).

\bibitem{Chen03}
Q. Chen, S. Chen, G.L. Eyink, 
``The joint cascade of energy and helicity in three-dimensional 
turbulence,'' 
{\it Phys. Fluids} {\bf 15}, 361-374 (2003).

\bibitem{Gomez04}
D.O. G\'omez and P.D. Mininni, 
``Understanding turbulence through numerical simulations,'' 
{\it Physica A} {\bf 342}, 69-75 (2004).

\bibitem{Brandenburg01}
A. Brandenburg, 
``The inverse cascade and nonlinear alpha-effect in simulations of 
isotropic helical hydromagnetic turbulence,'' 
{\it Astrophys. J.} {\bf 550}, 824-840 (2001).

\bibitem{Kazantsev67}
A.P. Kazanstev, 
``Enhancement of a magnetic field by a conducting fluid,'' 
{\it Sov. Phys. JETP} {\bf 26}, 1031-1034 (1968).

\end{thebibliography}
\end{document}